%% file: paper.tex
\documentclass[conference, a4paper, hidelinks]{IEEEtran}
\IEEEsettopmargin{t}{1.92cm}

\IEEEoverridecommandlockouts
\usepackage{cite}
\usepackage{amsmath,amssymb,amsfonts}
\usepackage{algorithmic}
\usepackage{graphicx}
\usepackage{textcomp}
\usepackage{xcolor}
\usepackage{soul}
\usepackage{todonotes}
\usepackage{dblfloatfix} 
\usepackage{bm}
\usepackage{upgreek}
\makeatletter
\let\MYcaption\@makecaption
\makeatother

\usepackage[font=footnotesize,labelformat=simple]{subcaption}

\makeatletter
\let\@makecaption\MYcaption
\makeatother
\usepackage{tikz}
\usepackage{pgfplots}
\usetikzlibrary{shapes,arrows,positioning,calc, matrix,spy,patterns}
\usepackage[nolist]{acronym}
\pgfplotsset{compat=1.18}

\usepackage{booktabs}
\usepackage{xfrac}
\usepackage{comment}
\usepackage{orcidlink}

\newcommand*{\figures}{figures}

\definecolor{cb-1}{HTML}{4477AA}
\definecolor{cb-2}{HTML}{EE6677}
\definecolor{cb-3}{HTML}{228833}
\definecolor{cb-4}{HTML}{CCBB44}
\definecolor{cb-5}{HTML}{66CCEE}
\definecolor{cb-6}{HTML}{AA3377}
\definecolor{cb-7}{HTML}{BBBBBB}

\pgfplotscreateplotcyclelist{cb list}{
	cb-1,cb-2,cb-3,cb-4,cb-5,cb-6,cb-7
}

\input{\makros/KIT_colors.tex}
\input{\makros/math.tex}
\input{\makros/tikz_makros.tex}

\def\BibTeX{{\rm B\kern-.05em{\sc i\kern-.025em b}\kern-.08em
    T\kern-.1667em\lower.7ex\hbox{E}\kern-.125emX}}

\begin{document}
\input{acronyms.tex}
\title{Precoding Design for Multi-User MIMO \\Joint Communications and Sensing
\thanks{This work has received funding 
	in part from the European Research Council
	(ERC) under the European Union’s Horizon 2020 research and innovation
	programme (grant agreement No. 101001899) and in part 
	from the German
	Federal Ministry of Research, Technology and Space (BMFTR) within the projects
	Open6GHub+ (grant agreement 16KIS2405) and KOMSENS-6G (grant agreement 16KISK123).}
}


\author{\IEEEauthorblockN{Charlotte Muth\,\orcidlink{0000-0002-0478-6085}, \IEEEmembership{Graduate Student Member, IEEE}, 
Shrinivas Chimmalgi\,\orcidlink{0000-0002-5868-3102},\\
and Laurent Schmalen\,\orcidlink{0000-0002-1459-9128}, \IEEEmembership{Fellow, IEEE}}


\IEEEauthorblockA{Communications Engineering Lab (CEL), Karlsruhe Institute of Technology (KIT)\\ 
		Hertzstr. 16, 76187 Karlsruhe, Germany, 
		Email: \texttt{first.last@kit.edu}\vspace*{-1ex}}
}


\maketitle

\begin{abstract}
We investigate precoding for \ac{MU} \ac{MIMO} \ac{JCAS} systems, taking into account the potential interference between sensing and communication channels. We derive indicators for the sensing and communication performance, i.e., the detection probability and the communication \ac{SINR} for general input signals. Our results show that the use of the communication signal for sensing can prevent a loss in communication performance if channel interference occurs, while the kurtosis of the transmit alphabet of the communication signal limits the sensing performance. We present simulation results of example setups.  
\end{abstract}



\input{\texfiles/content_nn_jcas_2.tex}

\bibliography{IEEEabrv,literature_short.bib}
\bibliographystyle{IEEEtran}
\begin{appendices}

\end{appendices}
\end{document}

%% file: makros/KIT_colors.tex
\definecolor{kit-green100}{rgb}{0,.59,.51}
\definecolor{kit-green70}{rgb}{.3,.71,.65}
\definecolor{kit-green50}{rgb}{.50,.79,.75}
\definecolor{kit-green30}{rgb}{.69,.87,.85}
\definecolor{kit-green15}{rgb}{.85,.93,.93}
\definecolor{KITgreen}{rgb}{0,.59,.51}

\definecolor{KITpalegreen}{RGB}{130,190,60}
\colorlet{kit-maigreen100}{KITpalegreen}
\colorlet{kit-maigreen70}{KITpalegreen!70}
\colorlet{kit-maigreen50}{KITpalegreen!50}
\colorlet{kit-maigreen30}{KITpalegreen!30}
\colorlet{kit-maigreen15}{KITpalegreen!15}

\definecolor{KITblue}{rgb}{.27,.39,.66}
\definecolor{kit-blue100}{rgb}{.27,.39,.67}
\definecolor{kit-blue70}{rgb}{.49,.57,.76}
\definecolor{kit-blue50}{rgb}{.64,.69,.83}
\definecolor{kit-blue30}{rgb}{.78,.82,.9}
\definecolor{kit-blue15}{rgb}{.89,.91,.95}

\definecolor{KITyellow}{rgb}{.98,.89,0}
\definecolor{kit-yellow100}{cmyk}{0,.05,1,0}
\definecolor{kit-yellow70}{cmyk}{0,.035,.7,0}
\definecolor{kit-yellow50}{cmyk}{0,.025,.5,0}
\definecolor{kit-yellow30}{cmyk}{0,.015,.3,0}
\definecolor{kit-yellow15}{cmyk}{0,.0075,.15,0}

\definecolor{KITorange}{rgb}{.87,.60,.10}
\definecolor{kit-orange100}{cmyk}{0,.45,1,0}
\definecolor{kit-orange70}{cmyk}{0,.315,.7,0}
\definecolor{kit-orange50}{cmyk}{0,.225,.5,0}
\definecolor{kit-orange30}{cmyk}{0,.135,.3,0}
\definecolor{kit-orange15}{cmyk}{0,.0675,.15,0}

\definecolor{KITred}{rgb}{.63,.13,.13}
\definecolor{kit-red100}{cmyk}{.25,1,1,0}
\definecolor{kit-red70}{cmyk}{.175,.7,.7,0}
\definecolor{kit-red50}{cmyk}{.125,.5,.5,0}
\definecolor{kit-red30}{cmyk}{.075,.3,.3,0}
\definecolor{kit-red15}{cmyk}{.0375,.15,.15,0}

\definecolor{KITpurple}{RGB}{160,0,120}
\colorlet{kit-purple100}{KITpurple}
\colorlet{kit-purple70}{KITpurple!70}
\colorlet{kit-purple50}{KITpurple!50}
\colorlet{kit-purple30}{KITpurple!30}
\colorlet{kit-purple15}{KITpurple!15}

\definecolor{KITcyanblue}{RGB}{80,170,230}
\colorlet{kit-cyanblue100}{KITcyanblue}
\colorlet{kit-cyanblue70}{KITcyanblue!70}
\colorlet{kit-cyanblue50}{KITcyanblue!50}
\colorlet{kit-cyanblue30}{KITcyanblue!30}
\colorlet{kit-cyanblue15}{KITcyanblue!15}

\definecolor{cb-1}{HTML}{4477AA}
\definecolor{cb-2}{HTML}{EE6677}
\definecolor{cb-3}{HTML}{228833}
\definecolor{cb-4}{HTML}{CCBB44}
\definecolor{cb-5}{HTML}{66CCEE}
\definecolor{cb-6}{HTML}{AA3377}
\definecolor{cb-7}{HTML}{BBBBBB}

\pgfplotscreateplotcyclelist{cb list}{
	cb-1,cb-2,cb-3,cb-4,cb-5,cb-6,cb-7
}

%% file: makros/math.tex
\let\j\relax
\newcommand{\j}{\mathrm{j}}
\let\Re\relax

\DeclareMathOperator{\Re}{Re}

\DeclareMathAlphabet{\mathbfsf}{\encodingdefault}{\sfdefault}{bx}{n}

\newcommand{\rscal}[1]{\mathsf{#1}}          

\newcommand{\vect}[1]{\mathbf{#1}}       
\newcommand{\rvect}[1]{\boldsymbol{\mathsf{#1}}} 

\newcommand{\mat}[1]{\mathbf{#1}}        
\newcommand{\rmat}[1]{\boldsymbol{\mathsf{#1}}}  

%% file: makros/tikz_makros.tex
\usepackage{ellipsis}
\usetikzlibrary{calc, patterns,fit,shapes}
\usetikzlibrary{decorations.pathreplacing,decorations.markings,shapes.geometric}
\usepgfplotslibrary{fillbetween}
\tikzset{naming/.style={align=center,font=\small}}
\tikzset{antenna/.style={insert path={-- coordinate (ant#1) ++(0,0.25) -- +(135:0.25) + (0,0) -- +(45:0.25)}}}
\tikzset{station/.style={naming,draw,shape=dart,shape border rotate=90, minimum width=10mm, minimum height=10mm,outer sep=0pt,inner sep=3pt}}
\tikzset{mobile/.style={naming,draw,shape=rectangle,minimum width=12mm,minimum height=6mm, outer sep=0pt,inner sep=3pt}}
\tikzset{radiation/.style={{decorate,decoration={expanding waves,angle=90,segment length=4pt}}}}
\usetikzlibrary{decorations.pathreplacing}

\tikzset{pics/MUE/.style args={#1}
{code={
\node[mobile,label={[inner ysep=+.3333em]\dots}] (box){};
\draw ([xshift=.25cm] box.south west) circle (4pt)
      ([xshift=-.25cm]box.south east) circle (4pt);

\fill ([xshift=.25cm] box.south west) circle (1pt)
      ([xshift=-.25cm]box.south east) circle (1pt);

\draw ([xshift=.25cm] box.north west) [antenna=1];
\draw ([xshift=-.25cm]box.north east) [antenna=2];
}}}

\tikzset{pics/UE/.style args={#1}
{code={
\node[mobile] (box) {#1};

\draw ([xshift=.25cm] box.south west) circle (4pt)
      ([xshift=-.25cm]box.south east) circle (4pt);

\fill ([xshift=.25cm] box.south west) circle (1pt)
      ([xshift=-.25cm]box.south east) circle (1pt);

\draw (box.north) [antenna=1];
}}}

\tikzset{pics/MBS/.style args={#1}{code={
\node[station] (-base) {#1};
\draw[line join=bevel] (-base.100) -- (-base.80) -- (-base.110) -- (-base.70) -- (-base.north west) -- (-base.north east);
\draw[line join=bevel] (-base.100) -- (-base.70) (-base.110) -- (-base.north east);
\node (-a1) at ([xshift=.3cm,yshift=.3cm] -base.north) {};
\node (-a2) at ([xshift=-.3cm,yshift=.3cm] -base.north) {};
\draw[line cap=rect] ([xshift=.3cm,yshift=.3pt] -base.north) [antenna=1];
\draw[line cap=rect] ([yshift=.3pt]ant1 |- -base.north) -- ([xshift=-.3cm,yshift=.3pt] -base.north) [antenna=2];
\node (-d) at ([yshift=1cm] -base.east) {\dots};

}}}

\tikzset{pics/BS/.style args={#1}{code={

\node[station] (base) {#1};

\draw[line join=bevel] (base.100) -- (base.80) -- (base.110) -- (base.70) -- (base.north west) -- (base.north east);
\draw[line join=bevel] (base.100) -- (base.70) (base.110) -- (base.north east);

\draw[line cap=rect] ([yshift=0pt]base.north) [antenna=1];

}}}

\tikzset{pics/car/.style args={#1}{code={
    \node (-base) at (4.2,2){};
    \draw[very thick,#1, rounded corners=0.18ex,fill=black!20!blue!20!white]  (2.7,1.8) -- ++(1,0.7) -- ++(1.6,0) -- ++(0.6,-0.7) -- (2.7,1.8);
        \draw[thick,#1, fill=#1,thick] (4.2,2.5) -- (5.3,2.5)  -- ++(0.2,-0.23) -- (4.4,2.27) -- (4.4,1.8) --(4.2,1.8) -- cycle;
      \draw[thick]  (4.2,1.8) -- (4.2,2.5);
      \fill[draw=#1,fill=#1,rounded corners=0.6ex,very thick] (1.5,.5) -- ++(0,1) -- ++(1.2,0.3) --  ++(3,0) -- ++(1,0) -- ++(0,-1.3) -- (1.5,.5) -- cycle;
      \draw[draw=black,fill=gray!50,thick] (2.75,.5) circle (.5);
      \draw[draw=black,fill=gray!50,thick] (5.5,.5) circle (.5);
      \draw[draw=black,fill=gray!80,semithick] (2.75,.5) circle (.4);
      \draw[draw=black,fill=gray!80,semithick] (5.5,.5) circle (.4);
      \filldraw[draw=#1, fill=KITorange, semithick] (2.1,1.3) -- ++(150:.5) arc (155:205:0.5) -- cycle;
  
}}}

\tikzset{
  pobl/.style={
    inner sep=0pt, outer sep=0pt, fill=#1,
  },
  pobl gron/.style n args={2}{
    pobl=#1, rounded corners=#2,
  },
  pics/person/.style n args={3}{
    code={
      \node (-corff) [pobl=#1, minimum width=.25*#2, minimum height=.375*#2, rotate=#3, pic actions] {};
      \node (-pen) [minimum width=.3*#2, circle, pobl=#1, outer sep=.01*#2, anchor=south, rotate=#3, pic actions] at (-corff.north) {};
      \node (-coes dde) [pobl gron={#1}{1pt}, anchor=north west, minimum width=.12125*#2, minimum height=.25*#2, rotate=#3, pic actions] at (-corff.south west) {};
      \node [pobl=#1, anchor=north, minimum width=.12125*#2, minimum height=.15*#2, rotate=#3, pic actions] at (-coes dde.north) {};
      \node (-coes chwith) [pobl gron={#1}{1pt}, anchor=north east, minimum width=.12125*#2, minimum height=.25*#2, rotate=#3, pic actions] at (-corff.south east) {};
      \node [pobl=#1, anchor=north, minimum width=.12125*#2, minimum height=.15*#2, rotate=#3, pic actions] at (-coes chwith.north) {};
      \node (-braich dde) [pobl gron={#1}{.75pt}, minimum width=.075*#2, minimum height=.325*#2, outer sep=.0064*#2, anchor=north west, rotate=#3, pic actions] at (-corff.north east)  {};
      \node [pobl=#1, minimum width=.05*#2, minimum height=.2*#2, outer sep=.0064*#2, anchor=north west, rotate=#3, pic actions] at (-corff.north east) {};
      \node (-braich chwith) [pobl gron={#1}{.75pt}, minimum width=.075*#2, minimum height=.325*#2, outer sep=.0064*#2, anchor=north east, rotate=#3, pic actions] at (-corff.north west) {};
      \node [pobl=#1, minimum width=.0375*#2, minimum height=.2*#2, outer sep=.0064*#2, anchor=north east, rotate=#3, pic actions] at (-corff.north west) {};
      \node (-fit person) [fit={(-pen.north) (-braich dde.east) (-coes chwith.south) (-braich chwith.west)}] {};
      \node (-pwy) [below=25pt of -fit person, every pin] {\tikzpictext};
      \draw [every pin edge] (-fit person) -- (-pwy);
    };
  };
}

\tikzset{pics/phone/.style args={#1}{code={
    \node (-base) at (1,1.75) {};
      \draw[draw=black,fill=black!80,rounded corners=0.2ex,very thick] (0,0) -- ++(0,3.5) -- ++(2,0) --  ++(0,-3.5) -- ++(-2,0) -- cycle;
      \fill[thick, fill=KITblue] (0.2,0.2) -- ++(0,3.1)  -- ++(1.6,0) -- ++(0,-3.1) -- ++(-1.6,0) -- cycle;
      \fill[fill=#1] (0.65,2) rectangle (0.75,2.1);
      \fill[fill=#1] (0.85,2) rectangle (0.95,2.3);
      \fill[fill=#1] (1.05,2) rectangle (1.15,2.5);
      \fill[fill=#1] (1.25,2) rectangle (1.35,2.7);
  
}}}

\newcommand{\hi}[2][KITyellow]{
	\ifmmode
	{
		 \mathchoice%
			{\colorbox{#1!50}{\textcolor{black}{$\displaystyle#2$}}}%
			{\colorbox{#1!50}{\textcolor{black}{$\textstyle#2$}}}
			{\colorbox{#1!50}{\textcolor{black}{$\scriptstyle#2$}}}
			{\colorbox{#1!50}{\textcolor{black}{$\scriptscriptstyle#2$}}}
	}
	\else
	{
		{\colorbox{#1!50}{\textcolor{black}{#2}}}
	}
	\fi
	}

%% file: acronyms.tex
\begin{acronym}[TROLOLO]
  \acro{ACM}{auto-correlation matrix}
  \acro{ADC}{analog to digital converter}
  \acro{AE}{autoencoder}
  \acro{ASK}{amplitude shift keying}
  \acro{AoA}{angle of arrival}
  \acro{AoD}{angle of departure}
  \acro{AWGN}{additive white Gaussian noise}
  \acro{BER}{bit error rate}
  \acro{BCE}{binary cross entropy}
  \acro{BICM}{bit-interleaved coded modulation}
  \acro{BMI}{bit-wise mutual information}
  \acro{BPSK}{binary phase shift keying}
  \acro{BP}{backpropagation}
  \acro{BSC}{binary symmetric channel}
  \acro{CAZAC}{constant amplitude zero autocorrelation waveform}
  \acro{CDF}{cumulative distribution function}
  \acro{CE}{cross entropy}
  \acro{CFAR}{constant false alarm rate}
  \acro{CLT}{central limit theorem}
  \acro{CNN}{concolutional neural network}
  \acro{CP}{cyclic prefix}
  \acro{CRB}{Cramér-Rao bound}
  \acro{CRC}{cyclic redundancy check}
  \acro{CSI}{channel state information}
  \acro{DFT}{discrete Fourier transform}
  \acro{DNN}{deep neural network}
  \acro{DoA}{degree of arrival}
  \acro{DOCSIS}{data over cable services}
  \acro{DPSK}{differential phase shift keying}
  \acro{DSL}{digital subscriber line}
  \acro{DSP}{digital signal processing}
  \acro{DTFT}{discrete-time Fourier transform}
  \acro{DVB}{digital video broadcasting}
  \acro{ELU}{exponential linear unit}
  \acro{ESPRIT}{Estimation of Signal Parameter via Rotational Invariance Techniques}
  \acro{FEC}{forward error correction}
  \acro{FFNN}{feed-forward neural network}
  \acro{FFT}{fast Fourier transform}
  \acro{FIR}{finite impulse response}
  \acro{GD}{gradient descent}
  \acro{GF}{Galois field}
  \acro{GMM}{Gaussian mixture model}
  \acro{GMI}{generalized mutual information}
  \acro{ICI}{inter-channel interference}
  \acro{IDE}{integrated development environment}
  \acro{IDFT}{inverse discrete Fourier transform}
  \acro{IFFT}{inverse fast Fourier transform}
  \acro{IIR}{infinite impulse response}
  \acro{ISI}{inter-symbol interference}
  \acro{JCAS}{joint communications and sensing}
  \acro{KKT}{Karush-Kuhn-Tucker}
  \acro{kldiv}{Kullback-Leibler divergence}
  \acro{LDPC}{low-density parity-check}
  \acro{LLR}{log-likelihood ratio}
  \acro{LTE}{long-term evolution}
  \acro{LTI}{linear time-invariant}
  \acro{LR}{logistic regression}
  \acro{MAC}{multiply-accumulate}
  \acro{MAP}{maximum a posteriori}
  \acro{MGF}{moment-generating function}
  \acro{MIMO}{multiple-input multiple-output}
  \acro{MLP}{multilayer perceptron}
  \acro{MLD}{maximum likelihood demapper}
  \acro{ML}{machine learning}
  \acro{MSE}{mean squared error}
  \acro{MLSE}{maximum-likelihood sequence estimation}
  \acro{MMSE}{miminum mean squared error}
  \acro{MU}{multi-user}
  \acro{NN}{neural network}
  \acro{NP}{Neyman-Pearson}
  \acro{OFDM}{orthogonal frequency-division multiplexing}
  \acro{OLA}{overlap-add}
  \acro{PAPR}{peak-to-average-power ratio}
  \acro{PDF}{probability density function}
  \acro{pmf}{probability mass function}
  \acro{PSD}{power spectral density}
  \acro{PSK}{phase shift keying}
  \acro{QAM}{quadrature amplitude modulation}
  \acro{QPSK}{quadrature phase shift keying}
  \acro{radar}{radio detection and ranging}
  \acro{RC}{raised cosine}
  \acro{RCS}{radar cross section}
  \acro{RMSE}{root mean squared error}
  \acro{RNN}{recurrent neural network}
  \acro{ROC}{receiver operating characteristic}
  \acro{ROM}{read-only memory}
  \acro{RRC}{root raised cosine}
  \acro{RV}{random variable}
  \acro{SER}{symbol error rate}
  \acro{SNR}{signal-to-noise ratio}
  \acro{SINR}{signal-to-interference-and-noise ratio}
  \acro{SPA}{sum-product algorithm}
  \acro{UE}{user equipment}
  \acro{ULA}{uniform linear array}
  \acro{VCS}{version control system}
  \acro{WLAN}{wireless local area network}
  \acro{WSS}{wide-sense stationary}
\end{acronym}

%% file: tex_files/content_nn_jcas_2.tex
\acresetall
\section{Introduction}
\Ac{MIMO} systems are fundamental to modern wireless communications, particularly in higher frequency bands where the increased path loss can be partially compensated through beamforming. \ac{MIMO} enables spatial multiplexing, i.e., using spatial diversity to transmit multiple data streams simultaneously to improve spectral efficiency.
Meanwhile, \ac{JCAS} has gained significant attention due to its potential to support new functionalities, such as warning vehicles at a crossing from potential collision risks. In \ac{JCAS}, sensing accuracy is closely linked to the available bandwidth, leading to  higher frequency bands being considered for sensing~\cite{Zhang_2022}.

A central challenge in \ac{JCAS} lies in the design of transmit signals that effectively support both communications and sensing.
While some research articles have proposed algorithms for precoding in \ac{MU}-\ac{MIMO} \ac{JCAS} systems~\cite{Wang2024,Dong2023,Nguyen25}, the question of transmit signal design is an ongoing field of research. In ~\cite{Wang2024, Ouyang23}, potential sensing targets are illuminated using phase-modulated communication signals, which yield maximum \ac{SNR} for object detection in single-input single-output systems. 
However, traditional communication systems often rely on
higher-order \ac{QAM} formats to increase throughput, which degrades sensing performance. To flexibly trade off communication and sensing performance, constellation shaping has been proposed~\cite{Geiger25} for \ac{JCAS}. 

While a \ac{JCAS} trade-off has been addressed for \ac{MIMO} in multiple studies, they tend to address performance metrics such as side-lobe levels or image \ac{SINR}~\cite{Keskin25} which are related to the average detection probability or the \ac{CRB} for parameter estimation~\cite{Hua2024}. Additionally, the communication signal streams use the same constellation and sensing is either done using communication signals or dedicated signals for sensing. We complement these results and methods by allowing communication signals of different constellations and directly optimizing the detection rate.


This work serves as a detailed analysis of the advantages and disadvantages of joint signaling or dedicated signals for sensing and investigates different scenarios. We obtain performance metrics, i.e., the target detection rate for a power detector and the communication \ac{SINR}. Lastly, we investigate how communications are affected by interference caused by sensing signals.

\acused{NN}
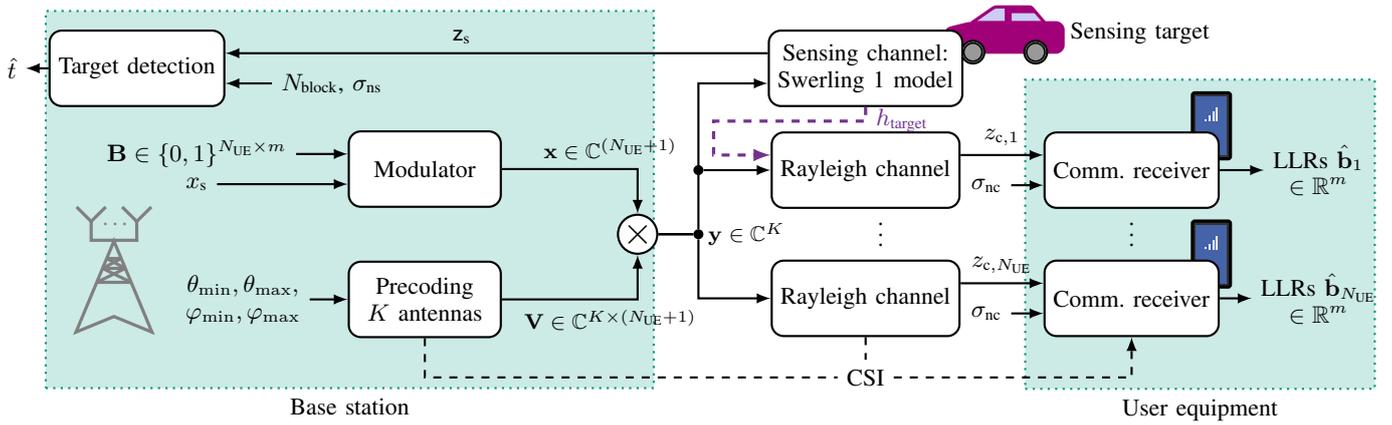
\begin{figure*}[t]
\vspace{2mm}
	\centerline{\hspace{-5mm}\input{\figures/flowgraph_alternative}}
	\vspace{-0.8cm}
	\caption{\ac{JCAS} system. The sensing target can become a reflector in the communication channel of \ac{UE}1, introducing interference $h_{\text{target}}$.}
	\label{fig:flowgraphtrain_mono}
	\vspace*{-0.4cm} 
\end{figure*}
\makeatletter
\AC@reset{NN}
\makeatother

\section{System Model}\label{sec:sysmodel}
We consider a single-carrier monostatic \ac{MU} \ac{MIMO} \ac{JCAS} system, where the transmitter and the sensing receiver are co-located at same base station with a \ac{ULA} with $K$ antennas and half-wavelength antenna spacing. Our goal is to detect a potential target based on the reflection of the transmitted signal at the target. Simultaneously, spatial multiplexing is used to communicate with $N_{\text{UE}}$ \acp{UE} with a single antenna located at different positions than the sensing target. The sensing target is located randomly at an azimuth angle of ${\theta} \in[\theta_{\min},\theta_{\max}]$. 
We consider multi-snapshot sensing with $N_{\text{block}}$ samples and consider for simplicity single target detection. The system block diagram is illustrated in Fig. \ref{fig:flowgraphtrain_mono}.
Throughout this paper, deterministic scalars, vectors, and matrices are denoted by $a$, $\vect{a}$, and $\mat{A}$, respectively; random scalars, vectors, and matrices are denoted by $\rscal{a}$, $\rvect{a}$, and $\rmat{A}$.

\subsection{Transmitter}
The communication transmitter generates the transmit bits for all \acp{UE}. We consider a transmit bit matrix $\mat{B} \nobreak \in \nobreak \{0,1\}^{N_{\text{UE}} \times m}$, with $m=4$ bits per symbol. The bits are mapped to symbols by the modulators, namely to $N_{\text{UE}}$ normalized 16-QAM signals for each \ac{UE}. Additionally, a constant-modulus signal ${x}_{\text{s}}$ is generated for sensing, e.g. unit power samples with uniformly distributed phase, and concatenated to a transmit vector $\vect{x}\in \mathbb{C}^{N_{\text{UE}}+1}$.
A linear precoder with weights $\vect{V}\in \mathbb{C}^{K \times (N_{\text{UE}}+1)}$ is used for beamforming.
The modulator and beamformer employ power normalization.
The transmit signal after linear precoding is
\begin{align}
    \vect{y} = \mat{V}\vect{x}.
\end{align}

\subsection{Channels}
A part of the radiated power is steered toward each \ac{UE} by the precoder while another part is steered toward the sensing area of interest.
The signal propagation from $K$ antennas towards an azimuth angle $\varphi$ is modeled with the spatial angle vector $\vect{a}_{\text{TX}}({\varphi}) \in \mathbb{C}^{K}$ given by
$
	 \vect{a}_{\text{TX}}(\varphi) = \nobreak \left(1,\text{e}^{\j\pi  \sin\varphi},\ldots, \text{e}^{\j\pi (K-1) \sin\varphi}\right)^\top.$
 For the communication part, the signal $\vect{y}$ experiences Rayleigh fading. 
Signals transmitted at various angles of departure (\acsp{AoD}) $\boldsymbol{\upvarphi}_u$ are received at the $u$th \ac{UE} due to multipath propagation, as reflections from surrounding surfaces cause the signal to arrive from multiple directions.
All reflections from $\boldsymbol{\upvarphi}_u= (\varphi_{1u}, \varphi_{2u}, \ldots )$ are assumed to arrive simultaneously at the \acp{UE} as spatial taps, i.e.
\begin{align}
 {z}_{\text{c},u} &=  
    \sum_{d=1}^{\dim(\boldsymbol{\upvarphi}_u)} \alpha_{\text{c},d} \vect{a}_{\text{TX}}({\varphi}_{du})^{\top} \mat{V} \vect{x} + {n}_{\text{c},u}\label{eq:comm-channel},
\end{align}
with fading coefficients ${{\alpha}_{\text{c},d} \sim \mathcal{CN}(0,\sigma_{\text{c},d}^2)}$, noise samples ${{n}_{\text{c},u}\nobreak \sim \nobreak \mathcal{CN}(0,\sigma_\text{nc}^2)}$ and channel $\tilde{\vect{h}}_{\text{c},du}:= \nobreak \alpha_{\text{c},d} \vect{a}_{\text{TX}}(\boldsymbol{\upvarphi}_{du})^{\top} \mat{V}$.
To stress the importance of channel interference, we let the sensing target act as a reflective element within the channel of \ac{UE}1 while additional user interference can be caused by precoding.
We can rephrase \eqref{eq:comm-channel} for \ac{UE}1 as
\begin{align}
    z_{\text{c},1} &= 
    \sum_{d=1}^{\dim(\boldsymbol{\upvarphi}_{u1})}  \tilde{\vect{h}}_{\text{c},d1} \vect{x} + {n_{\text{c}}}
= \left( \sum_{u=1}^{N_{\text{UE}}+1} h_u x_{u} \right)+ {n_{\text{c}}}\notag \\
    &= {h}_{1}x_{1} + \underbrace{\left( \sum_{u=2}^{N_{\text{UE}}} h_u x_{u} \right)}_{\text{user interference}} + {h}_{\text{target}}\underbrace{x_{N_{\text{UE}}+1}}_{=x_{\text{s}}} + \underbrace{{n_{\text{c}}}}_{\text{noise}},
\end{align}
with $\vect{h} = (h_1, \ldots, h_{N_{\text{UE}}+1})= \sum_d \tilde{\vect{h}}_{\text{c},d1}$ including precoding, fading, and additional multipath interference. We have rewritten the sensing interference as $h_{\text{target}}={h}_{N_{\text{UE}}+1}$.


With $t\in \nobreak \{0,1\}$ indicating the presence of the target at an angle $\theta$, we express the sensing signal reflected from said target in the monostatic setup as
\begin{align}
	\rvect{{z}}_{\text{s}} = t\,  \vect{a}_{\text{RX}}({\theta}) \mathsf{a} \vect{a}_{\text{TX}}({\theta})^\top \vect{y}
      + {\rvect{n}}_{\text{s}},\label{eq:sens-channel}
\end{align}
with the radar target following a Swerling-1 model~\cite{Swerling1960} with random variable ${\rscal{a}} \sim \mathcal{CN}(0,\sigma_{\text{s}}^2)$ representing the radar cross section and path loss and noise $\rvect{n}_{\text{s}} = (\rscal{n}_{\text{s},1}, \ldots, \rscal{n}_{\text{s},K})$ with $\rscal{n}_{\text{s},k} \sim \nobreak \mathcal{CN}(0,\sigma_{\text{ns}}^2)$ and $k\in \{ 1, \ldots, K\}$. The spatial angle vectors relate as $\vect{a}_{\text{RX}}({\theta})=\vect{a}_{\text{TX}}({\theta})$, with $\theta$ being the \ac{AoD} and \ac{AoA} of the target. The radial velocity of the target is assumed to be zero, so no Doppler shift occurs.

\subsection{Communication Receivers}
The goal of the communication receiver is to recover the transmitted bits based on the received signal. We assume that channel estimation has already been performed at the communication receiver and that the precoding matrix $\mat{V}$ is known. 
Therefore, full \ac{CSI} is available at each \ac{UE} to perform (MMSE) equalization.
The demodulator outputs \acp{LLR} $\hat{\vect{b}} \in \mathbb{R}^{m}$ that can be used as input to a soft decision channel decoder. 

\subsection{Sensing Receiver}
We process $N_{\text{block}}$ realizations of \eqref{eq:sens-channel} for detection. With a Swerling-1 model, we model scan-to-scan deviations of the \ac{RCS}, which manifest as a change in $\rscal{a}$ during the observation window. Each realization also results in its own noise samples and transmit symbols. Therefore, we rewrite the samples of \eqref{eq:sens-channel} for a time instance $\ell \in \{1,2,\ldots, N_{\text{block}}\}$ and antenna element $k$, with \(c_\ell =\nobreak\vect{a}_{\text{TX}}(\theta)^{\top}\mat{y}_{\ell}\) as
\begin{align}
    \rscal{z}_{\text{s},k\ell} = t\,  \text{e}^{\j\pi(k-1)  \sin\theta} \rscal{a}_{\ell} c_\ell
      + \rscal{n}_{\text{s},k\ell}.\label{eq:sens-channel-l}
\end{align}
Although prior knowledge of the transmit signal or approximate target positions may be available in a monostatic sensing scenario, in this work we assume that the transmit signal is unknown at the sensing receiver as well as the possible \ac{AoA} of a target. This assumption provides a meaningful lower bound on performance when such information is available, while enabling closed-form expressions and analysis for the resulting detector.

A detector with false alarm probability $P_{\text{f}}$ based on the \ac{NP} criterion~\cite{Trees2002} is given by
\begin{align}
    \frac{2}{\sigma_{\text{ns}}^2}\sum_{\ell=1}^{N_{\text{block}}} \sum_{k=1}^{K} |\rscal{z}_{\text{s},k\ell}|^2 \quad\mathop{\gtreqless}_{\hat{t}=0}^{\hat{t}=1}\quad Q_{\chi^2_{2KN_{\text{block}}}}(1-P_{\text{f}}),\label{eq:detector}
\end{align}
with $Q_{\chi^2_{2KN_{\text{block}}}}(\cdot)$ denoting the quantile function of a $\chi^2$ distribution  with ${2KN_{\text{block}}}$ degrees of freedom. Thus, if the left-hand side of \eqref{eq:detector} is higher than the detection threshold, a target is detected.
Knowing the noise statistics, we can implement this power detector as a \ac{CFAR} detector.
This choice of detector enables us to analyze the detection statistics in closed form.
For $t=0$ only noise is present, yielding a $\chi^2$ distribution. For $t=1$ the detector input is influenced by beamforming and transmit signals, making the analysis more involved.

\section{Performance Indicators for Sensing and Communications}\label{sec:performance}

\subsection{Detection Probability}\label{sec:detection-prob}
The detector can be formulated in closed form using \eqref{eq:detector}.
We can drop the receive phase shift $\text{e}^{\j\pi(k-1)  \sin\theta}$ from~\eqref{eq:sens-channel-l}, as only the received power is of interest. A single sample for the detector can be reformulated as
\begin{align}
    |\rscal{z}_{\text{s},k\ell}|^2 = |t c_{\ell}\rscal{a}_{\ell} + \rscal{n}_{\text{s},k\ell}|^2.
\end{align}
In the following, we derive a closed-form expression of the detector statistics for $t=1$ to further analyze the impact of precoding on the detection probability.

\textbf{Derivation:} 
$\rscal{n}_{\text{s},k \ell}$ and $\rscal{a}_{\ell}$ are independent circularly symmetric complex Gaussian random variables with zero mean and variances
\[
\mathbb{E}\{|\rscal{a}_\ell|^2\} = \sigma_{\text{s}}^2, \quad \mathbb{E}\{|\rscal{n}_{\text{s},k \ell}|^2\} = \sigma_{\text{ns}}^2.
\]
Let \(c_\ell =\sum_{u=1}^{N_{\text{UE}}}\sqrt{\beta_u}x_{u\ell}\) be fixed constants for now, with $\sqrt{\beta_u}=\vect{a}_{\text{TX}}(\theta)^{\top}(\mat{v})_u$ denoting the effects of beamforming. 
We keep the beamforming gain $\beta_u$ constant by assuming approximately uniform illumination of sensing targets, otherwise the detection probability would be dependent on the \ac{AoA} of the target. We want to derive the distribution of
\begin{equation}
    \tilde{\rscal{z}}_{\text{s}} = \sum_{\ell=1}^{N_{\text{block}}}  \sum_{k=1}^K|\rscal{z}_{\text{s},k\ell}|^2= \sum_{\ell=1}^{N_{\text{block}}}  \sum_{k=1}^K |c_\ell\rscal{a}_\ell+\rscal{n}_{\text{s},k\ell}|^2
\end{equation}
to analyze the detector behavior. As $c_\ell$, $\rscal{a}_\ell$ and $\rscal{n}_{\text{s},k\ell}$ are independently drawn from their respective distributions for each time sample $\ell$, we focus on a single $\ell$ where we are interested in the distribution of $\sum_{k=1}^K |c_\ell\rscal{a}_\ell+\rscal{n}_{\text{s},k\ell}|^2$
with $\rscal{a}_\ell$ being independent of $k$.  
The received samples for the $K$ antennas are components of a multivariate distribution of a variable $\rvect{q}=[c_\ell\rscal{a}_\ell+\rscal{n}_{\text{s},1\ell}, c_\ell\rscal{a}_\ell+\rscal{n}_{\text{s},2\ell},\ldots,c_\ell\rscal{a}_\ell+\rscal{n}_{\text{s},K\ell}]^{\top}$. Then $\rvect{q}$ is a complex circular Gaussian with zero mean and covariance matrix
\begin{equation}
    \mat{\Sigma}_\ell=\sigma_{\text{ns}}^2 \bm{I}+|c_\ell|^2\sigma_{\text{s}}^2\bm{1}_{K\times K},
\end{equation}
with all-one matrix $\bm{1}_{K \times K} \in {1}^{K \times K}$.
The covariance matrix $\mat{\Sigma}_\ell$ is Hermitian positive-semi-definite and, hence, can be decomposed using the singular value decomposition as 
\begin{equation}
    \mat{\Sigma}_\ell = \bm{U}\bm{\Lambda}\mat{U}^H, \quad \bm{\Lambda}=\mathrm{diag}(\lambda_1,\cdots,\lambda_K).
\end{equation}
Let $\rvect{r}=\mat{U}^H\rvect{q}$ and, thus, since $\mat{U}$ is unitary, $\rvect{q}=\mat{U}\rvect{r}$. Then $\rvect{r}\sim\mathcal{CN}(\bm{0},\bm{\Lambda})$, and all components $\rscal{r}_k\sim \mathcal{CN}(0,\lambda_k)$ are independent Gaussians.
We can write
\begin{align}
    \sum_{k=1}^K |c_\ell\rscal{a}_\ell+\rscal{n}_{\text{s},k\ell}|^2 &=\rvect{q}^H\rvect{q}
    =(\mat{U}\rvect{r})^H (\mat{U}\rvect{r})
    =\sum_{k=1}^K |\rscal{r}_k|^2   . 
\end{align}
The variances $\lambda_k$ of $\rscal{r}_k$ are the eigenvalues of the covariance matrix $\mat{\Sigma}_\ell$. As $\mat{\Sigma}_\ell$ is the sum of a diagonal matrix and a rank-1 matrix, the eigenvalues are known in closed form 
\begin{align}
    \lambda_1=\sigma_{\text{ns}}^2+K|c_\ell|^2\sigma_{\text{s}}^2,\quad
    \lambda_2=\cdots=\lambda_K=\sigma_{\text{ns}}^2.
\end{align}

Knowing that $|\rscal{r}_k|^2$ follows an exponential distribution with rate parameter equal to $1/\lambda_k$ 
and that the \ac{MGF} of the sum of independent variables is given by the product of the \acp{MGF} of the variables~\cite{Pho2019}, the \ac{MGF} of $\tilde{\rscal{z}}_{\text{s},\ell}=\sum_{k=1}^K |c_\ell\rscal{a}_\ell+\rscal{n}_{\text{s},k\ell}|^2$ is given by
\begin{equation}
M_{\tilde{\rscal{z}}_{\text{s},\ell}}(s) = \frac{1}{1-(\sigma_{\text{ns}}^2+K|c_\ell|^2\sigma_{\text{s}}^2)s}\left(\frac{1}{1-\sigma_{\text{ns}}^2s}\right)^{K-1}.
\end{equation}

Given the independent time observations, the \ac{MGF} of $\tilde{\rscal{z}}_{\text{s}}= \nobreak \sum_{\ell=1}^{N_\text{block}}\tilde{\rscal{z}}_{\text{s},\ell}$ is given by
\begin{equation}
M_{\tilde{\rscal{z}}_{\text{s}}}\!(s) = \left(\!\frac{1}{1-\sigma_{\text{ns}}^2s}\!\right)^{\!(K-1)N_\text{block}}\cdot\prod_{\ell=1}^{N_\text{block}}\!\!\frac{1}{1-(\sigma_{\text{ns}}^2+K|c_\ell|^2\sigma_{\text{s}}^2)s}.
\end{equation}

The \ac{PDF} $f_{\tilde{\rscal{z}}_{\text{s}}}(z)$ can be obtained from the \ac{MGF} through the inverse Laplace transform
\begin{align}
    f_{\tilde{\rscal{z}}_{\text{s}}}(z) &= \mathcal{L}^{-1}\{M_{\tilde{\rscal{z}}_{\text{s}}}(-s)\}\label{eq:mgf_pdf}.
\end{align}

Maximizing the detection probability, given by
\begin{align}
    P_{\text{d}} = \int_{\frac{\sigma_{\text{ns}}^2}{2} Q_{\chi^2_{2KN_{\text{block}}}}(1-P_{\text{f}})}^{\infty} f_{\tilde{\rscal{z}}_{\text{s}}}(z)\, \mathrm{d}z \label{eq:Pd}
\end{align}
requires either increasing the mean of $f_{\tilde{\rscal{z}}_{\text{s}}}(z)$ or decreasing its variance.
We calculate the mean and variance of the \ac{PDF} in the presence of a target ($t=1$) through the \ac{MGF}, resulting in:
\begin{align}
    \mu =& \frac{\mathrm{d}}{\mathrm{d}s}M_{\tilde{\rscal{z}}_{\text{s}}}(s=0) = KN_{\text{block}}\sigma_{\text{ns}}^2 +K\sigma_{\text{s}}^2 \left(\sum_{\ell=1}^{N_{\text{block}}} |c_\ell|^2\right)\notag \\
    =& KN_{\text{block}}\sigma_{\text{ns}}^2 +K\sigma_{\text{s}}^2 \sum_{\ell=1}^{N_{\text{block}}}\left|\sum_{u=1}^{N_{\text{UE}}+1}\sqrt{\beta_u}x_{u\ell}\right|^2\label{eq:mu1},\, \text{and}
\end{align}
\begin{align}
    \sigma_{{\tilde{\rscal{z}}_{\text{s}}}}^2 =& \frac{\mathrm{d}^2}{\mathrm{d}s^2}M_{\tilde{\rscal{z}}_{\mathrm{s}}}(s=0)-\mu^2\notag\\
    =& N_{\text{block}}K\sigma_{\text{ns}}^4+K^2\sigma_{\text{s}}^4\sum_{\ell=1}^{N_{\text{block}}}\left|\sum_{u=1}^{N_{\text{UE}}+1}\sqrt{\beta_u}x_{u\ell}\right|^4 \notag\\+ &2K\sigma_{\text{ns}}^2\sigma_{\text{s}}^2 \sum_{\ell=1}^{N_{\text{block}}}\left|\sum_{u=1}^{N_{\text{UE}}+1}\sqrt{\beta_u}x_{u\ell}\right|^2\label{eq:var1}.
\end{align}
We now consider the transmit symbols as random variables $\rvect{x} \in \mathbb{C}^{N_{\text{UE}}+1} \sim P_{\rvect{x}}$.
For sufficiently large $N_{\text{block}}$, we can apply the law of large numbers over $N_{\text{block}}$ and constrain the launch power using $\mathbb{E}_{P_{\rvect{{x}}}}\{ |{\rvect{{x}}}|^2 \}= 1$ (i) and $
\sum_{u=1}^{N_{\text{UE}}+1} |\beta_u| = 1$ (ii) with $\mathbb{E}_{P_{\rvect{{x}}}} \{{\rvect{{x}}}\}=0$ (iii). 
If the signals for different \acp{UE} are uncorrelated, we can approximate:
\begin{align}
    \lambda &:= \mathbb{E}_{P_{\rvect{{x}}}} \left\{ \left|\sum_{u=1}^{N_{\text{UE}}+1}\sqrt{\beta_u} \rscal{x}_{u}\right|^2\right\}\notag \\
    &=  |\beta_1|\mathbb{E}_{P_{\rvect{{x}}}}\left\{|{\rscal{x}}_{1}|^2\right\} + \ldots + \mathbb{E}_{P_{\rvect{{x}}}}\left\{2\Re\{\sqrt{\beta_1\beta_2^*}{\rscal{x}}_{1}{\rscal{x}}_{2}^*\}\right\}\notag \\
    &= \sum_{u=1}^{N_{\text{UE}}+1}|\beta_u| = 1\\
    \tilde{\kappa} &:=  \mathbb{E}_{P_{\rvect{{x}}}}\left\{\left| \sum_{u=1}^{N_{\text{UE}}+1}\sqrt{\beta_u} {\rscal{x}}_{u}\right|^4 \right\}\notag \\
    &= |\beta_1|^2 \underbrace{\mathbb{E}_{P_{\rvect{{x}}}}\left\{|{\rscal{x}}_{1}|^4\right\}}_{\kappa_1} + \ldots 
    +4|\beta_1||\beta_2| + \ldots \notag \\
    &\stackrel{\text{(ii)}}{=} \sum_{u=1}^{N_{\text{UE}}+1} |\beta_u|^2 \kappa_u + 2|\beta_u|(1-|\beta_u|),\label{eq:kappa}
\end{align}
with kurtosis $\kappa_u$ of $\rscal{x}_u$. 
Then, \eqref{eq:mu1} and \eqref{eq:var1} can be approximated as
\begin{align*}
    \mu &\approx KN_{\text{block}}(\sigma_{\text{ns}}^2 +\sigma_{\text{s}}^2) \\
    \sigma_{{\tilde{\rscal{z}}_{\text{s}}}}^2 &\approx N_{\text{block}}K(\sigma_{\text{ns}}^4+K\sigma_{\text{s}}^4\tilde{\kappa} + 2\sigma_{\text{ns}}^2\sigma_{\text{s}}^2).
\end{align*}

\subsection{Influence of Precoding}
As stated before, the detection performance can be maximized through increasing $\mu$ or decreasing $\sigma_{{\tilde{\rscal{z}}_{\text{s}}}}^2$. 
The detection probability within the constraints (i) -- (iii) is affected by the transmit signal and beamforming only through $\tilde{\kappa}$ given by~\eqref{eq:kappa}.
The larger the variance of the distribution, the lower a potential detection rate becomes. For fixed constellations $P_{\rvect{{{x}}}}$ and ${\rvect{{{x}}}}$, we compute the Hessian matrix of $\tilde{\kappa} (|\beta_1|,\ldots, |\beta_{N_{\text{UE}+1}}|)$ as
\begin{align}
    \mat{H}_{\tilde{\kappa}} = \begin{pmatrix}
         2\kappa_1-4 &\cdots & 0\\
        \vdots &\ddots & \vdots\\
        0 & \cdots & 2\kappa_{N_{\text{UE}+1}}-4 \\
        \end{pmatrix}.
\end{align}
For any vector $\vect{w} \in \mathbb{R}^{N_{\text{UE}+1}}$, 
\begin{align}
    \vect{w}^\top \mat{H}_{\tilde{\kappa}} \vect{w} = \sum_{u=1}^{N_{\text{UE}+1}} w_u^2 (2\kappa_u-4) \leq 0, 
\end{align}
as $1 \leq \kappa_u \leq 2$ for realistic constellation kurtosis. 
If $\kappa_u=2$ for all $u$, which applies for Gaussian distributed constellation points, $\tilde{\kappa}$ is affine and the choice of $|\beta_u|$ becomes irrelevant, leading to identical performance.
In all other cases, $\mat{H}_{\tilde{\kappa}}$ is negative definite and $\tilde{\kappa}$ is strictly concave with respect to $|\beta_u|$.
Then, the minima of $\tilde{\kappa}$ can be found at the boundaries of the feasible set, with the global minimum at 
\begin{align}
    |\beta_u|=\begin{cases}
    1, &u= \arg\min_u \kappa_u \\ 
    0, &\text{otherwise}.
\end{cases}
\end{align}
In words, the detection probability is maximized if the target is illuminated only by the signal with the lowest $\kappa_u$.

\subsection{Communications \ac{SINR}}
The \ac{SINR} for communication, using normalized constellations is given by
\begin{align}
    \text{SINR}_\text{c} = \frac{|h_1|^2}{\sum_{u=2}^{N_{\text{UE}}}|h_u|^2 +|h_{\text{target}}|^2 + \sigma_{\text{nc}}^2}\label{eq:SINRc-sep}.
\end{align}
To analyze the effect of channel interference, we examine the case where the signal for \ac{UE}1 illuminates the sensing area resulting in signal $x_{\text{s}}=x_{1}$.
Then, the \ac{SINR} is
\begin{align}
    \text{SINR}_\text{cs} = \frac{|h_1+h_{\text{target}}|^2}{\sum_{u=2}^{N_{\text{UE}}}|h_u|^2 + \sigma_{\text{nc}}^2}.
\end{align}
Depending on the phase difference $\phi = \angle h_1 - \angle h_{\text{target}}$ which we assume to be uniformly distributed, the resulting \ac{SINR} can be higher or lower compared to \eqref{eq:SINRc-sep}. We obtain the mean \ac{SINR} as
\begin{align}
    \overline{\text{SINR}}_\text{cs} &=  \frac{\int_{-\pi}^{\pi} \left(|h_1|^2+|h_{\text{target}}|^2 + 2\Re\{|h_1h_{\text{target}}| \mathrm{e}^{\j\phi}\} \right)\mathrm{d}\phi}{2\pi \left(\sum_{u=1}^{N_{\text{UE}}}|h_u|^2 + \sigma_{\text{nc}}^2 \right)}\notag \\
    &= \frac{|h_1|^2+|h_{\text{target}}|^2}{\sum_{u=2}^{N_{\text{UE}}}|h_u|^2 + \sigma_{\text{nc}}^2}.
\end{align}
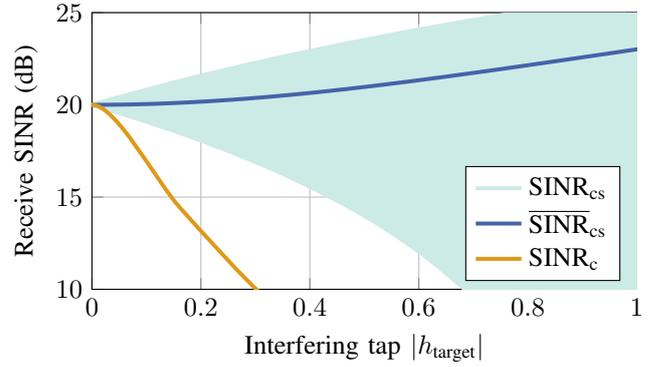
\begin{figure}
    \centering
    \begin{tikzpicture}
	\begin{axis}[grid=both,
			xmin=0, xmax=1,
			ymin= 10, ymax=25,
			xlabel={Interfering tap $|h_{\text{target}}|$},
			ylabel={Receive SINR (dB)},
                legend entries={$\text{SINR}_\text{cs}$,,,$\overline{\text{SINR}}_\text{cs}$,$\text{SINR}_\text{c}$}, 
			legend cell align={left},
                legend pos=south east,
                height=0.6\columnwidth,
                width=\columnwidth]
	\addplot[name path=A,smooth,black,mark=none,
	line width=1.5pt,
	samples=63,
	color=KITgreen!20, label={$\text{SINR}_\text{cs}$}]  {10*log10((1+x)*(1+x)/0.01)};
	\addplot+[name path=B, smooth,black,mark=none,
	line width=1.5pt,
	samples=140,
	color=KITgreen!20]  {10*log10((1-x)*(1-x)/0.01)};
	\addplot+[KITgreen!20] fill between [of=A and B];
	\addplot+[ smooth,mark=none,
line width=1.5pt,
samples=140,
color=KITblue, label={$\Bar{\text{SINR}}_\text{cs}$}]  {10*log10(((1)^2+x^2)/0.01)}; 
	
	\addplot+[smooth,mark=none,
	line width=1.5pt,
	samples=63,
	color=KITorange, label={$\text{SINR}_\text{c}$}]  {10*log10((1)^2/(0.01+x^2))};
	\end{axis}
	\end{tikzpicture}
    \vspace{-4mm}
\caption{Communication \ac{SINR} with $\text{SNR}_{\text{c}} =20\,$dB and $h_1=1$.}
\label{fig:comm-snr}
\vspace{-6mm}
\end{figure}

Fig.~\ref{fig:comm-snr} displays the possible \ac{SINR} range and average \ac{SINR} resulting from separate signaling for sensing or reuse of a communication signal. Even if additional interference from a sensing signal is not phase-coherent, $\overline{\text{SINR}}_\text{cs}$ actually rises with increased interference $|h_{\text{target}}|$. In a scenario where interference is caused by reflections in the environment, we cannot expect the signals to have the same phase, so this is an encouraging result from a communications perspective.


\section{Simulation Setup}

We train an \ac{AE} framework similar to~\cite{Muth25Access} with precoding and demodulation implemented as \acp{NN}\footnote{Code at \url{https://github.com/frozenhairdryer/JCAS-MIMO-precoding}}.
A weight parameter $w_{\text{s}} \in [0,1]$ controls the impact of the detection task, resulting in the loss
\begin{align}
    L = (1-w_{\text{s}})L_{\text{comm}} + w_{\text{s}}L_{\text{detect}} \label{eq:loss},
\end{align}
with the loss terms given by
\begin{align}
   {L}_{\text{comm}} =
-\sum_{u=1}^{N_{\text{UE}}}\log \left(\sum_{i=1}^{m} 1-\mathrm{H}(\mathsf{b}_{iu}||\mathsf{\hat{b}}_{iu})\right) ,\,
   L_{\text{detect}} = \mathrm{H}(\mathsf{t}||\hat{\mathsf{t}}),\notag
   \label{eq:lossalt}
\end{align}
with $\mathrm{H}(\cdot||\cdot)$ denoting the binary cross entropy between two binary distributions.
The communication loss is a reformulated utility function, leading to maximum fairness~\cite{Castaneda2017}.

Minimizing $L_{\text{detect}}$ corresponds to maximizing the detection probability \eqref{eq:Pd}. As shown in Sec.~\ref{sec:detection-prob}, the detection probability is maximized when $\tilde{\kappa}$ is minimized. As $\tilde{\kappa}$ is concave with respect to resource allocation, gradient descent will lead to illumination of the sensing area by whichever signal is initially strongest. Therefore, it is particularly sensitive to the initialization. To incentivize convergence to the global minimum, we initialize $\mat{V}$ such that $\tilde{\kappa}$ is maximum.

\section{Results and Discussion}\label{sec:sims}
We simulate two \acp{UE} at \acp{AoA} of $\varphi =\{50^\circ,70^\circ\}$. A potential radar target is located in a range $\theta \in [-20^\circ,20^\circ]$.
For transmission and sensing, we simulate a \ac{ULA} with $K=16$ and we consider an observation window of $N_{\text{block}}=15$. For the sensing receiver, we set the false alarm rate to $P_{\text{f}}=\nobreak10^{-2}$.
The signal power is $\sigma_{\text{c}}^2=1$ and noise power $\sigma_{\text{nc}}^2=0.01$ for both \acp{UE} and we train a respective decoder. We set $w_{\text{s}}=0.2$ and vary the amount of interference for targets with $\theta \in [10^{\circ},20^{\circ}]$ to \ac{UE}1 by varying $h_{\text{target}}/c \sim \mathcal{N}(0,{\sigma}_{\text{c},\text{target}}^2)$.
We compare three different cases for the sensing receiver:
\begin{itemize}
    \item $\tilde{\rscal{z}}_{\mathrm{s},\text{CM}}$ describes sensing using only a constant modulus phase-modulated signal $x_{\text{s}}$ for sensing.
    \item $\tilde{\rscal{z}}_{\mathrm{s},\text{QAM}}$ describes sensing using the communication 16-QAM signals $x_1, \ldots, x_{N_{\text{UE}}}$.
    \item $\tilde{\rscal{z}}_{\mathrm{s},\text{QAM}+\text{CM}}$ allows for a combination of a dedicated sensing signal $x_{\text{s}}$ and communication signals $x_1, \ldots, x_{N_{\text{UE}}}$.
\end{itemize}


\subsection{Detector Statistics}
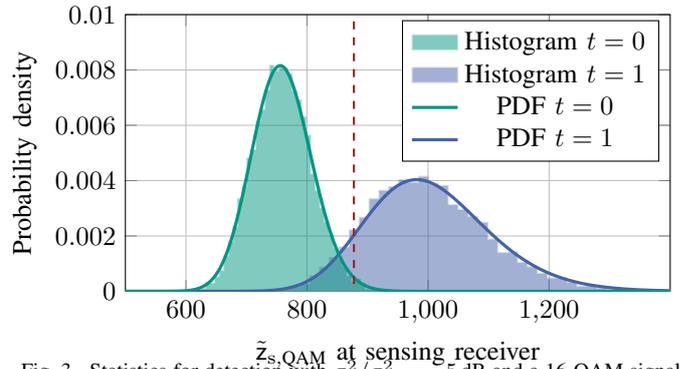
\begin{figure}
\hspace{-3.5mm}
    \begin{tikzpicture}
\begin{axis}[ xlabel={$\tilde{\rscal{z}}_{\mathrm{s},\text{QAM}}$ at sensing receiver}, ylabel=Probability density,
xmin=500, xmax=1400,width=\columnwidth, height=0.6\columnwidth,
ymin=0, ymax=0.01,
scaled ticks=false,
tick label style={/pgf/number format/fixed},
/pgf/number format/precision=3,
grid=major,
xtick={600, 800, 1000, 1200},
]
\addlegendimage{area legend,fill=KITgreen, draw=KITgreen!50, opacity=0.5}\addlegendentry{Histogram $t=0$}
\addlegendimage{area legend,fill=KITblue, draw=KITblue!50, opacity=0.5}\addlegendentry{Histogram $t=1$}
\addlegendimage{KITgreen, very thick}\addlegendentry{PDF $t=0$}
\addlegendimage{KITblue, very thick}\addlegendentry{PDF $t=1$}
    \addplot [
        const plot,
        fill=KITblue,
        opacity=0.5,
        draw=KITblue!50,
    ] table[x=v1,y=bin1] {\figures/hist_data2.txt}
        \closedcycle;
    \addplot [
        const plot,
        fill=KITgreen,
        opacity=0.5,
        draw=KITgreen!50,
    ] table[x=v2,y=bin2] {\figures/hist_data2.txt}
        \closedcycle;
    \addplot[KITblue,very thick, no marks] table[x=p, y=t1] {\figures/pd_mgf2.txt};
    \addplot[KITgreen,very thick, no marks] table[x=p, y=t0] {\figures/pd_mgf2.txt};
    \addplot[KITred, dashed, thick, no marks] coordinates {(877.5418147829952,0.1) (877.5418147829952,0)};
\end{axis}
\end{tikzpicture}
\vspace{-6mm}
\caption{Statistics for detection with $\sigma_{\text{s}}^2/\sigma_{\text{ns}}^2=-5\,$dB and a 16-QAM signal. The dashed line indicates the detection threshold for $P_{\text{f}}=0.01$.}
\label{fig:hist-pdf}
\vspace{-4mm}
\end{figure}

In Fig.~\ref{fig:hist-pdf}, the empirical histograms, the \acp{PDF} obtained from \eqref{eq:mgf_pdf} and the detection threshold are shown for the case of no target ($t=0$) and a target illuminated by a 16-QAM signal ($t=1$). We obtained the \acp{PDF} by generating $50$ realizations for each $|c_l|$, calculating the mean of $M_{\tilde{\rscal{z}}_{\mathrm{s}}}(s)$ and then applying the numerical inverse Laplace transform. As the Laplace transform is a linear transform, this is equivalent to the mean \ac{PDF}. The histograms and \acp{PDF} match very well. For smaller noise power $\sigma_{\text{ns}}^2$, both \acp{PDF} shift to the left and become more narrow. A larger signal power $\sigma_{\text{s}}^2$ results in the \ac{PDF} for $t=1$ being shifted to the right and widened.

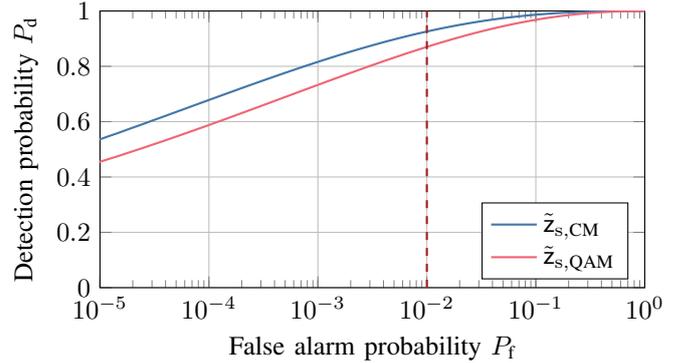
\begin{figure}
	\begin{tikzpicture}
		\begin{semilogxaxis}[xlabel=False alarm probability $P_{\text{f}}$,         
                ylabel={Detection probability $P_{\text{d}}$},
			 grid=major,
            legend entries= {$\tilde{\rscal{z}}_{\mathrm{s},\text{CM}}$, $\tilde{\rscal{z}}_{\mathrm{s},\text{QAM}}$},
			 legend cell align={left},
			 legend pos=south east,
			 xmin=1e-5,xmax=1,
		ymin=0,ymax=1,
                 width=\columnwidth,
                 height=0.6\columnwidth,
			]
            \addplot+[mark=none, color=cb-1, thick] table [x expr = 1-\thisrow{pf-qam}, y expr= 1- \thisrow{pd-psk}]
			{\figures/roc_plot.txt};
            \addplot+[mark=none, color=cb-2, thick] table [x expr = 1-\thisrow{pf-qam}, y expr= 1-\thisrow{pd-qam}]
			{\figures/roc_plot.txt};

            \addplot[KITred, dashed, thick, no marks] coordinates {(0.01,0) (0.01,1)};
\end{semilogxaxis}
\end{tikzpicture}
\vspace{-6mm}
\caption{Detection probability with $\sigma_{\text{s}}^2/\sigma_{\text{ns}}^2=-5\,$dB for phase-modulated and 16-\ac{QAM} sensing input. The dashed line indicates the operating point of $P_{\text{f}}=0.01$ used for further evaluation.}
\label{fig:mimo-pb-cs}
\vspace{-6mm}
\end{figure}

In the case of independent channels for sensing and communication, we get optimal sensing performance if a dedicated constant-modulus signal is used for sensing.
In Fig.~\ref{fig:mimo-pb-cs}, $P_{\text{d}}$ for changing $P_{\text{f}}$ based on~\eqref{eq:mgf_pdf} is shown. The detector performs consistently better under $\tilde{\rscal{z}}_{\mathrm{s},\text{CM}}$, as the signal has a lower kurtosis $\kappa$ than $\tilde{\rscal{z}}_{\mathrm{s},\text{QAM}}$. For very high false alarm rates this gap can vanish. 
\begin{figure}
\vspace{-0.65cm}
	\begin{tikzpicture}
		\begin{axis}[
			xlabel=Angle (deg), ylabel=Beamforming gain (dB),
			legend entries={UE1,UE2, Sensing}, 
            legend style={font=\footnotesize},
			legend cell align={left},
			legend pos=north west,
			xmin=-90,xmax=90,
			ymin=-30,ymax=10,
			axis line style=thick,
                width=\columnwidth,
                height=0.6\columnwidth,
                extra x ticks={-25, 25, 75,-75},
                tick align=inside,
                grid=major,
			]
            \fill[fill=KITblue, fill opacity=0.1] (axis cs: -20,-40) rectangle (axis cs: 20,10);
            \fill[fill=KITpurple, fill opacity=0.1] (axis cs: 10,-40) rectangle (axis cs: 20,10);
            
            \addplot [color=cb-2, thick] table [x expr= {\thisrow{angles}}, y expr = {10*log10(\thisrow{Ephi0})}]
		{\figures/mimo_beta/beampattern0.txt};
        
            \addplot [color=cb-3, thick] table [x expr= {\thisrow{angles}}, y expr = {10*log10(\thisrow{Ephi1})}]
		{\figures/mimo_beta/beampattern0.txt};
            \addplot [color=cb-1, thick] table [x expr= {\thisrow{angles}}, y expr = {10*log10(\thisrow{Ephi2})}]
		{\figures/mimo_beta/beampattern0.txt};
            \draw [black] (axis cs:0,7) node {\small Sensing};
            \node[KITpurple, rotate=-90] (ue1) at (axis cs:54.5,6) {\small UE1};
            \node[KITpurple, rotate=-90] at (axis cs:74.5,6) {\small UE2};

            \addplot [color=cb-2, thick, dotted] table [x expr= {\thisrow{angles}}, y expr = {10*log10(\thisrow{Ephi0})}]
		{\figures/mimo_beta_-5db_cb/beampattern8.txt};
            \addplot [color=cb-3, thick, dotted] table [x expr= {\thisrow{angles}}, y expr = {10*log10(\thisrow{Ephi1})}]
		{\figures/mimo_beta_-5db_cb/beampattern8.txt};

            \addplot [color=cb-2, thick,dashed] table [x expr= {\thisrow{angles}}, y expr = {10*log10(\thisrow{Ephi0})}]
		{\figures/mimo_beta/beampattern24.txt};
            \addplot [color=cb-3, thick,dashed] table [x expr= {\thisrow{angles}}, y expr = {10*log10(\thisrow{Ephi1})}]
		{\figures/mimo_beta/beampattern24.txt};
        \addplot [color=cb-1, thick,dashed] table [x expr= {\thisrow{angles}}, y expr = {10*log10(\thisrow{Ephi2})}]
		{\figures/mimo_beta/beampattern24.txt};
            
            \addplot [mark=none, KITpurple,thick] coordinates {(50, -50) (50, 30)};
            \addplot [mark=none, KITpurple,thick] coordinates {(70, -50) (70, 30)};

            \draw[->, KITpurple!50!KITblue, -latex] (axis cs:15,6) -- (ue1) node[midway, below, yshift=-0mm] 
            {\small{${h}_{\text{target}}$}};

            \addplot[mark=none,black, thick] coordinates {(0,-80) (1,-80)};\label{plot:b0}
            \addplot[mark=none,black, dotted, thick] coordinates {(0,-80) (1,-80)};\label{plot:b1}
            \addplot[mark=none,black, dashed, thick] coordinates {(0,-80) (1,-80)};\label{plot:b12}
\end{axis}
\end{tikzpicture}
\vspace{-4mm}
\caption{Beam patterns for two \acp{UE} using 16-QAM  with example scenarios \protect\ref{plot:b0} $\tilde{\rscal{z}}_{\mathrm{s},\text{CM}}$, \protect\ref{plot:b12} $\tilde{\rscal{z}}_{\mathrm{s},\text{QAM}+\text{CM}}$, and \protect\ref{plot:b1} $\tilde{\rscal{z}}_{\mathrm{s},\text{QAM}}$.}
\label{fig:mimo-beam-beta}
\vspace{-1mm}
\end{figure}
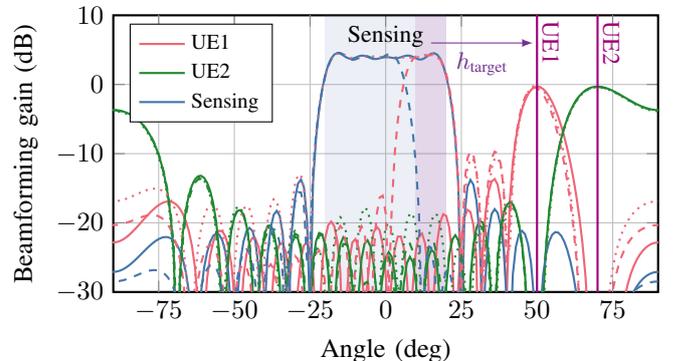

\subsection{Precoding Evaluation}
Trained beam patterns for the different approaches are shown in Fig.~\ref{fig:mimo-beam-beta}. For $\tilde{\rscal{z}}_{\mathrm{s},\text{CM}}$ resulting from ${\sigma}_{\text{c},\text{target}}=0$, the whole area of interest for sensing is illuminated by the constant modulus sensing signal. For ${\sigma}_{\text{c},\text{target}}\geq0.08$, the beam pattern switches to $\tilde{\rscal{z}}_{\mathrm{s},\text{QAM}+\text{CM}}$, as the section causing interference is is illuminated with the signal of \ac{UE}1. For $\tilde{\rscal{z}}_{\mathrm{s},\text{QAM}}$, the signal of \ac{UE}1 illuminates the whole sensing area.


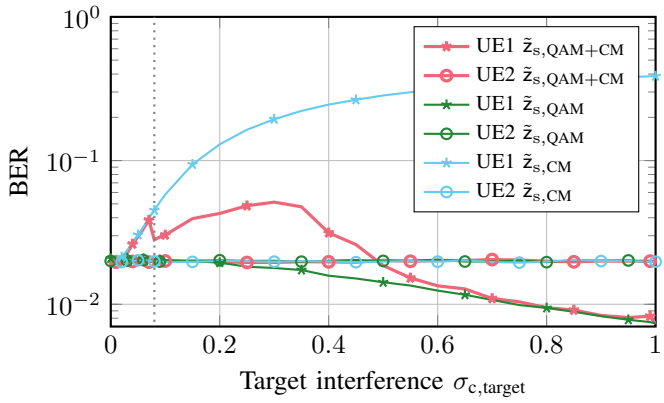
\begin{figure}
	\begin{tikzpicture}
		\begin{semilogyaxis}[
			xlabel=Target interference ${\sigma}_{\text{c},\text{target}}$,         ylabel=BER,
			grid=major,
           legend entries= {UE1 $\tilde{\rscal{z}}_{\mathrm{s},\text{QAM}+\text{CM}}$, UE2 $\tilde{\rscal{z}}_{\mathrm{s},\text{QAM}+\text{CM}}$, UE1 $\tilde{\rscal{z}}_{\mathrm{s},\text{QAM}}$, UE2 $\tilde{\rscal{z}}_{\mathrm{s},\text{QAM}}$, UE1 $\tilde{\rscal{z}}_{\mathrm{s},\text{CM}}$, UE2 $\tilde{\rscal{z}}_{\mathrm{s},\text{CM}}$},
			legend cell align={left},
			legend pos=north east,
			xmin=0,xmax=1,
			ymin=0.007,ymax=1,
			axis line style=thick,
                width=\columnwidth,
                height=0.65\columnwidth,
                legend style={name=leg},
                legend style={font=\footnotesize}
			]
                \addplot[mark=star, color=cb-2, very thick, mark repeat=3,mark phase=2] table [x expr = \thisrow{beta}, y = BER0]
			{\figures/betasweep_BER.txt};
                \addplot[mark=o, color=cb-2, very thick,  mark repeat=3,mark phase=2] table [x expr = \thisrow{beta}, y = BER1]
			{\figures/betasweep_BER.txt};

            \addplot[mark=star, color=cb-3, thick, mark repeat=3,mark phase=1] table [x expr = \thisrow{beta}, y = BER0]
			{\figures/betasweep_BER_comm.txt};
                \addplot[mark=o, color=cb-3, thick,  mark repeat=3,mark phase=1] table [x expr = \thisrow{beta}, y = BER1]
			{\figures/betasweep_BER_comm.txt};

            \addplot[mark=star, color=cb-5, thick, mark repeat=3,mark phase=3] table [x expr = \thisrow{beta}, y = BER0]
			{\figures/betasweep_BER_sensonly.txt};
                \addplot[mark=o, color=cb-5, thick,  mark repeat=3,mark phase=3] table [x expr = \thisrow{beta}, y = BER1]
			{\figures/betasweep_BER_sensonly.txt};
        \addplot[gray, dotted, thick] coordinates {(0.08,0.001) (0.08,1)};
\end{semilogyaxis}
\end{tikzpicture}
\vspace{-6mm}
\caption{BER at $\text{SNR}_{\text{c}}=20\,$dB for 2 UEs using 16-QAM for modulation and the beam patterns of Fig.~\ref{fig:mimo-beam-beta} for different target interference at \ac{UE}1.
}
\label{fig:mimo-ber}
\vspace{-4mm}
\end{figure}
Concerning communication performance, the \ac{BER} for $t=1$ and the target being at angle $\theta \in [10^{\circ},20^{\circ}]$ is shown in Fig.~\ref{fig:mimo-ber}. 
The \ac{BER} for \ac{UE}2 remains constant. For \ac{UE}1, the \ac{BER} associated with $\tilde{\rscal{z}}_{\mathrm{s},\text{QAM}+\text{CM}}$ rises first because the whole (for ${\sigma}_{\text{c},\text{target}}<0.08$) or part of the interfering area is illuminated by the constant-modulus signal, reducing the \ac{SINR}. With ${\sigma}_{\text{c},\text{target}}>0.6$, the target is illuminated by the 16-QAM signal only, approaching the performance of $\tilde{\rscal{z}}_{\mathrm{s},\text{QAM}}$.
For $\tilde{\rscal{z}}_{\mathrm{s},\text{QAM}}$, the \ac{BER} for \ac{UE}1 decreases slowly. 
If the whole sensing area is consistently illuminated with a constant-modulus signal in scenario $\tilde{\rscal{z}}_{\mathrm{s},\text{CM}}$, the \ac{BER} of \ac{UE}1 increases significantly with increasing $\sigma_{\text{c},\text{target}}$. These findings match the expectations from Fig.~\ref{fig:comm-snr}. 

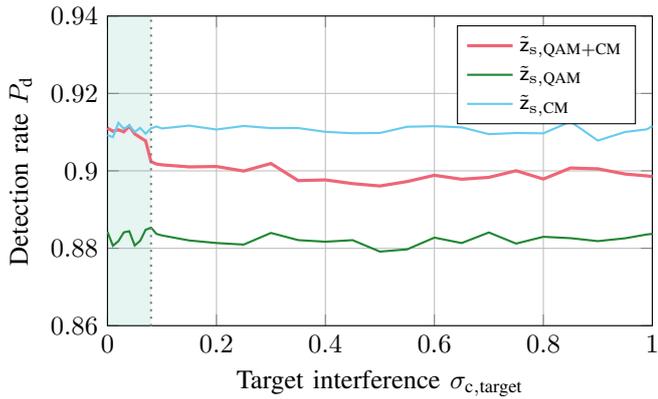
\begin{figure}
	\begin{tikzpicture}
		\begin{axis}[xlabel=Target interference ${\sigma}_{\text{c},\text{target}}$,         
                ylabel={Detection rate $P_{\text{d}}$},
			 grid=major,
            legend entries= {$\tilde{\rscal{z}}_{\mathrm{s},\text{QAM}+\text{CM}}$, $\tilde{\rscal{z}}_{\mathrm{s},\text{QAM}}$, $\tilde{\rscal{z}}_{\mathrm{s},\text{CM}}$},
			 legend cell align={left},
			 legend pos=north east,
			 xmin=0,xmax=1,
		ymin=0.86,ymax=0.94,
                 width=\columnwidth,
                 height=0.65\columnwidth,
                 legend style={font=\footnotesize}
			]
            
            \addplot+[mark=none, color=cb-2, very thick] table [x expr = \thisrow{beta}, y = detect_prob_29]
			{\figures/mimo_beta_-5db/betasweep_Pd.txt};
            
            \addplot+[mark=none, color=cb-3, thick] table [x expr = \thisrow{beta}, y = detect_prob_29]
			{\figures/betasweep_Pd_comm.txt};

            \addplot+[mark=none, color=cb-5, thick] table [x expr = \thisrow{beta}, y = detect_prob_29]
			{\figures/betasweep_Pd_sensonly.txt};

            \addplot [name path=A,domain=0:0.08]{1};
 
            \addplot [name path=B,domain=0:0.08,samples=2]{0.3};
 
            \addplot [KITgreen, fill opacity=0.1, ] fill between [of=A and B];
            
            \label{plot:sb}
            \addplot[gray, dotted, thick] coordinates {(0.08,0.3) (0.08,1)};
            %
\end{axis}
\end{tikzpicture}
\vspace{-6mm}
\caption{Detection rate with varied interference. The interfering sensing area is illuminated by the sensing signal in~\protect\ref{plot:sb} and by a beam of the communication signal of \ac{UE}1 otherwise. 
}
\label{fig:mimo-pb-beta}
\vspace{-6mm}
\end{figure}
Finally, we evaluate the detection rate based on ${\sigma}_{\text{c},\text{target}}$ in Fig.~\ref{fig:mimo-pb-beta}. 
If only $\tilde{\rscal{z}}_{\mathrm{s},\text{QAM}}$ or $\tilde{\rscal{z}}_{\mathrm{s},\text{CM}}$ are used, the detection rate remains constant independently of ${\sigma}_{\text{c},\text{target}}$, representing reference points for the achievable detection rate.
Using $\tilde{\rscal{z}}_{\mathrm{s},\text{QAM}+\text{CM}}$, we clearly see a decrease of the detection rate when the illumination source of the interfering area switches to the communication signal at ${\sigma}_{\text{c},\text{target}}=0.08$. The detection rate decreases slightly further and stabilizes around $P_{\text{d}}\approx 0.9$. 

To summarize, if significant interference occurs it is vital to perform sensing of spatially interfering components with a communication signal.

\section{Conclusion}\label{sec:concl}
\acused{ESPRIT}
In this paper, we provide guidance on how to design the transmit signal for \ac{MIMO} \ac{JCAS} systems based on the interference between sensing and communication channels. Areas of interest for sensing can be divided into sections that cause user interference for communication users and non-interfering sections and be handled separately as follows:
 If there is no interference, using a constant-modulus signal for sensing leads to optimal target detection without influencing communication performance.
If interference is observed, we encounter a trade-off. Using a communication signal for sensing aids in communication performance, but depending on the kurtosis of the used communication signal the sensing performance is limited.
Using the sum of two or more different signals for sensing is not recommended, as it leads to worse target detection than using a single signal.
The kurtosis of transmit signals being a mayor trade-off parameter for \ac{JCAS} remains consistent with the literature.


%% file: figures/flowgraph_alternative.tex
\tikzset{block/.style={rectangle, thick, draw, minimum width=2cm, minimum height=1cm, rounded corners=1.6mm},font=\small,align=center}
\tikzset{crossing/.style={circle, scale=0.4, fill=black}}
\vspace{0.5cm}
\pgfdeclarelayer{background}
\pgfdeclarelayer{foreground}
\pgfsetlayers{background,main,foreground}
\begin{tikzpicture}[node distance=1.5, text=black, >=latex]
	\node[] (start) {$\mat{B} \in \{0,1\}^{N_{\text{UE}} \times m}$};
    \node[] (start2) at ($(start)+(0,-0.4)$){$x_{\text{s}}$};
    
    \node[block, node distance=0.5cm, fill=white] (enc) at ($(start)+(3,-0.2)$) {Modulator};
	
	\node[block, node distance=0.7, below = of enc, fill=white] (beam){Precoding \\ $K$ antennas};
	\node[node distance=0.5cm, left = of beam] (theta){$\theta_{\min},\theta_{\max},$\\$ \varphi_{\min},\varphi_{\max}$};
	
	\node[draw,circle,inner sep=0, radius=0.4cm,fill=white, thick] (mix) at ($0.5*(beam)+0.5*(enc)+(2.8,0)$) {\Large{$\times$}} ;
	\node[] (mixtext) at (mix) {};  
	\node[block, fill=white] (radch) at ($(mix) + (3,2.2)$) {Sensing channel:\\ Swerling 1 model};
	\node[coordinate, node distance=5, left=of radch] (ac) {$\text{Corr}(\mat{z_{\text{s}}},\mat{z_{\text{s}}})$};
	\node[block] (commch) at ($(mix) + (3,-0.85)$) {Rayleigh channel};
    \node[block] (commch2) at ($(mix) + (3,0.85)$) {Rayleigh channel};
	\node[crossing] (c1) at ($(mix)+(0.8,0)$){};
        \node[crossing] (c3) at ($(commch2)-(2.2,0)$) {};
    	
	\node[block, fill=white] (detect) at ($(ac) - (3.3,0)$) {Target detection};

	\node[node distance=0.3cm, left = of detect] (llr) {$\hat{t}$};
	
	\node[block, fill=white] (demod) at ($(commch) + (3.5,0)$) {Comm. receiver};

    \node[block, fill=white] (demod2) at ($(commch2) + (3.5,0)$) {Comm. receiver};
	\node[node distance=0.3cm, below=of commch] (csi)  {CSI};
	\node[node distance=0.3cm] (bits) at ($(demod.east)-(-1.3,0)$) {LLRs $\hat{\vect{b}}_{N_{\text{UE}}}$\\$ \in \mathbb{R}^{m}$};
    \node[node distance=0.3cm] (bits2) at ($(demod2.east)-(-1.3,0)$) {LLRs $\hat{\vect{b}}_1$\\$ \in \mathbb{R}^{m}$};
        \node[] (i2) at ($(detect.east)+(1.4,-0.2)$){$N_{\text{block}}$, $\sigma_{\text{ns}}$};
        \node[] (i3) at ($(demod.west)+(-0.75,-0.2)$){$\sigma_{\text{nc}}$};
        \node[] (i4) at ($(demod2.west)+(-0.75,-0.2)$){$\sigma_{\text{nc}}$};
        \node[] (i5) at ($0.5*(commch)+0.5*(commch2)+(0.2,0.1)$){$\vdots$};
        \node[] (i6) at ($0.5*(demod)+0.5*(demod2)+(0,0.1)$){$\vdots$};

	\draw[thick, ->] (start) -- ($(enc.west)+(0,0.2)$);
    \draw[thick, ->] (start2) -- ($(enc.west)+(0,-0.2)$);
	\draw[thick, ->] (theta) -- (beam);
	\draw[thick, ->] (beam) -| (mix) node[pos=0.4,below] {\footnotesize $\mat{V} \in \mathbb{C}^{K \times(N_{\text{UE}}+1)}$};
	\draw[thick, ->] (enc) -| (mix) node[pos=0.4,above] {\footnotesize $\vect{x} \in \mathbb{C}^{(N_{\text{UE}}+1)}$};
	\draw[thick, ->] (mix) -- (c1) node[right] {\footnotesize $\vect{y}\in \mathbb{C}^{K}$} |-  ($(radch.west)-(0,0.2)$);
        \draw[thick, ->] (c3) -- (commch2);
	\draw[thick, ->] (mix) -- (c1) |- (commch);
	\draw[thick,->] ($(radch.west)+(0,0.2)$) -- +(-0.915,0) -- ($(detect.east)+(0,0.2)$)  node[midway,above]{$\vect{\mathsf{z}}_{\text{s}}$};
	\draw[thick, ->] (detect) -- (llr);
        \draw[thick, ->] (i2) -- ($(detect.east)-(0,0.2)$);
        \draw[thick, ->] (i3) -- ($(demod.west)-(0,0.2)$);
        \draw[thick, ->] (i4) -- ($(demod2.west)-(0,0.2)$);
	\draw[thick, ->, dashed] (commch) -- (csi) -| (demod);
	\draw[thick, ->] ($(commch.east)+(0,0.2)$) -- ($(demod.west)+(0,0.2)$) node[midway,above]{${{z}}_{\text{c},N_{\text{UE}}}$};
        \draw[thick, ->] ($(commch2.east)+(0,0.2)$) -- ($(demod2.west)+(0,0.2)$) node[midway,above]{${{z}}_{\text{c},1}$};
	\draw[thick, ->] ($(demod.east)-(0,0)$) -- (bits);
        \draw[thick, ->] ($(demod2.east)-(0,0)$) -- (bits2);
	\draw[thick, dashed] (beam.south) |- (csi);
    \draw[very thick, dashed, KITpurple!50!KITblue, ->] (radch.south) |- ($(radch)-(2,0.7)$) node[near start, right, yshift=-0.08cm] {${h}_{\text{target}}$} |- ($(commch2.west)+(0,0.2)$);

 \begin{pgfonlayer}{background}
\node[draw, minimum size=2cm, rectangle,minimum width = 8cm, minimum height=5cm, KITgreen, fill=KITgreen!20,thick, dotted,
	label=south:{Base station}] (A) at ($(ac)+(-0.5,-1.74)$) {};
 \pic[ node distance=0.5cm,left = of theta, scale=1, gray, very thick] (bs) {MBS};
 \pic[ scale=0.3, KITpurple, very thick] (car) at ($(radch.north east)+(-0.7,-0.45)$) {car};
 \node[label=east:{Sensing target}] (labcar) at ($(car-base)+(0.6,-0.2)$) {};
 \node[draw, minimum size=2cm, rectangle,minimum width = 4.6cm, minimum height=4.1cm, KITgreen, fill=KITgreen!20,thick, dotted,
	label=south:{User equipment}] (A) at ($(demod)+(0.9,0.85)$) {};
 \pic[thick, fill=white, scale=0.25] (UE1) at ($(demod)+(0.8,0.15)$) {phone};
 \pic[thick, fill=white, scale=0.25] (UE2) at ($(demod2)+(0.8,0.15)$) {phone};
\end{pgfonlayer}
\end{tikzpicture}